\documentclass{elsart4}

\usepackage{amssymb}

\usepackage{graphics}
\usepackage{graphicx}
\usepackage{epsfig}
\usepackage{amssymb}
\journal{Physica B}

\begin{document}

\begin{frontmatter}

\title{Anderson model out of equilibrium: conductance and Kondo temperature}

\author[label1]{L. Tosi} 
\author[label2]{P. Roura-Bas}
\author[label2]{A. M. Llois}
\author[label1]{A. A. Aligia}
\address[label1]{Centro At\'omico Bariloche and Instituto Balseiro, Comisi\'on Nacional de
Energ\'{\i}a At\'omica, 8400 Bariloche, Argentina}
\address[label2]{Dpto de F\'{\i}sica, Centro At\'{o}mico Constituyentes, Comisi\'{o}n 
Nacional de Energ\'{\i}a At\'{o}mica, Buenos Aires, Argentina}

\corauth[cor1]{Tel: 54-11 67727102; FAX: 54-11 67727121; roura@tandar.cnea.gov.ar}

\begin{abstract}

We calculate conductance through a quantum dot weakly coupled to metallic contacts
by means of Keldysh out of equilibrium formalism. We model the quantum dot with
the SU(2) Anderson model and consider the limit of infinite Coulomb repulsion. We
solve the interacting system with the numerical diagrammatic Non-Crossing
Approximation (NCA). We calculate the conductance
as a function of temperature and gate voltage, from differential conductance
(dI/dV) curves. We discuss these results in comparison with those from the linear
response approach which can be performed directly in equilibrium conditions.
Comparison shows that out of equilibrium results are in good agreement
with the ones from linear response supporting reliability to the method
employed. The discussion becomes relevant when dealing with general transport
models through interacting regions. We also analyze the evolution of 
the curve of conductance vs gate voltage with temperature. While at high temperatures 
the conductance is peaked when the Fermi energy coincides with the energy of the localized level, 
it presents a plateau 
for low temperatures as a consequence of Kondo effect. 
We discuss different ways to determine Kondo's
temperature and compare the values obtained in and out of equilibrium.
\end{abstract}

\end{frontmatter}

%%%%%%%%%%%%%%%%%%%%%%%%%%%%%%%%%%%%%%%%%%%%%%%%%%%%%%%%%%%%%%%%%%%%%%%%
\section{Introduction}
Since the first observation of the Kondo effect in semiconducting 
quantum dots (QD)\cite{goldhaber} the study of transport through 
nanoscopic devices has inspired a rich variety of
experimental and theoretical works. Nowadays, measurements of transport properties in such systems, 
such as current versus bias voltage and conductance, are the main focus of the experiments due to the interesting and 
unusual features observed \cite{roch,parks}. 

The behavior of the conductance at different temperatures and for different 
gate voltages has been studied in very general systems, including those 
showing strong correlations. While at equilibrium almost exact numerical methods have been 
developed for the theoretical treatment of this problems (numerical 
renormalization group (NRG)\cite{nrg} or exact diagonalization 
(ED)\cite{ed}), the ones for non-equilibrium conditions are still in progress. Among them, the Scattering Bethe Ansatz (SBA)\cite{SBA} and the Time dependent Density Matrix Renormalization Group (t-DMRG)\cite{t-DMRG} are promising. 

In this work we study the transport properties of an interacting QD 
using the Non-Crossing Approximation (NCA) in its 
non-equilibrium\cite{win} and equilibrium\cite{bickers} versions. We 
consider mandatory the comparison beetwen both schemes to support 
reliability to the more general procedure dealing with an out of equilibrium 
calculation. We discuss the results for conductance as function of 
bias and gate voltage, and moreover, the dependence of transport 
properties with temperature.

%%%%%%%%%%%%%%%%%%%%%%%%%%%%%%%%%%%%%%%%%%%%%%%%%%%%%%%%%%%%%%%%%%%%%%%%
\section{Model}
We study the transport properties through a quantum dot 
weak-coupled to metallic contacts and describe the system with the Anderson model,  
\begin{eqnarray}
H &=& \sum_{k\sigma\nu} \epsilon_{k\sigma\nu} c^{\dag}_{k\sigma\nu}
c_{k\sigma\nu} +\sum_{\sigma} E_{d} d^{\dag}_{\sigma}d_{\sigma} + U
d^{\dag}_{\uparrow}d_{\uparrow}d^{\dag}_{\downarrow}d_{\downarrow} \nonumber \\
&+& \sum_{k\nu\sigma} V_{k\nu\sigma} c^{\dag}_{k\nu\sigma}d_{\sigma} + H.c.,
\label{eq:ham_gral} 
\end{eqnarray}
where $c_{k\sigma\nu}$($c^{\dagger}_{k\sigma\nu}$) is the destruction 
(creation) operator of an electron with momentum $k$, spin $\sigma$ 
and lead $\nu=L$ (left) or $R$ (right), and 
$d_{\sigma}$ ($d^{\dagger}_{\sigma}$) destroys (creates) an electron 
in the quantum dot. 

The non-interacting conduction electrons in the leads are treated as being in thermal 
and chemical equilibrium with their reservoirs, thus 
$\epsilon_{k\sigma\nu} = \epsilon_{k\sigma} - \mu_{\nu}$, 
allowing for different chemical potentials $\mu_{\nu}$ in each of 
them. For the central region, we consider a spin degenerate localized level with energy $E_d$ and
Coulomb repulsion $U$. 
The leads and the dot are 
connected by means of hybridizations $V_{k\sigma\nu}$.

The physical quantity  accessible in 
transport measurements is the current.
As shown by Meir and Wingreen\cite{meir}, the current through a system 
described by the Hamiltonian Eq.(\ref{eq:ham_gral}) is given by
\begin{eqnarray}
I &=& \frac{2\pi e}{h} Tr\int d\omega  (\Gamma^{L} 
f_{L}(\omega) - \Gamma^{R} f_R(\omega) )\rho_{d}(\omega)
\nonumber \\
&+& \frac{2\pi e}{h}Tr \int d\omega  (\Gamma^L -\Gamma^R )iG_{d}^{<}(\omega)/2\pi, 
\label{eq:current}
\end{eqnarray}
where $\Gamma^{\nu}_{\sigma} = 2\pi \sum_{k} |V_{k\sigma\nu}|^2 
\delta(\omega - \epsilon_{k\sigma\nu}) $ is the hybridization 
function and $f_L$($f_R$) is the Fermi function for 
the conduction electrons of the left(right) lead. The functions 
$\rho_{d}(\omega)$ and $G_{d}^{<}(\omega)$ represent the spectral density 
and the lesser Green function of the central region respectively. 
The calculation of such Green functions must be done in presence of the leads
and is a non-equilibrium problem 
which might be handled within Keldysh formalism. 

From the results of different applied bias voltages, differential conductance 
$dI/dV$ curves can be 
obtained by differentiation of Eq. (\ref{eq:current}). The value at zero bias $dI/dV |_0=G(T)$ represent (the usual equilibrium 
conductance) at a given temperature. A simplified analytical expression for $G(T)$ can be obtained under the condition of  proportional couplings, $\Gamma^{L}_{\sigma} = \alpha \Gamma^{R}_{\sigma}$ \cite{win}
\begin{equation}
G(T) = 4\pi \frac{e^2}{h} \int d\omega\left( -\frac{\partial 
f}{\partial \omega}\right) \Gamma(\omega) 
\rho_{d}(\omega),   
\label{eq:conduct}
\end{equation}
where $\Gamma_{\sigma} = \Gamma^{L}_{\sigma}\Gamma^{R}_{\sigma}/(\Gamma^{L}_{\sigma} + 
\Gamma^{R}_{\sigma})$ is the effective hybridization and 
$\rho_{d}(\omega )$ is calculated in equilibrium. Thus, in contrast to Eq. (\ref{eq:current}) ony equilibrium quantities enter 
Eq. (\ref{eq:conduct}).

%%%%%%%%%%%%%%%%%%%%%%%%%%%%%%%%%%%%%%%%%%%%%%%%%%%%%%%%%%%%%%%%%%%%%%%%
\section{Results}

For the treatment of the model in and out of equilibrium, we 
use the diagrammatic NCA technique\cite{win,kroha,roura1}. 
We consider the case of infinite repulsion, $U\rightarrow \infty$. 
It must be noted that different sets of self-consistent equations have 
to be solved in order to compute the equilibrium and non-equilibrium
Green's functions. Therefore, computing the current from Eqs (\ref{eq:current}) and (\ref{eq:conduct})
allows us to test the validity of the methods used. 

\begin{figure}
\includegraphics[width=0.45\textwidth]{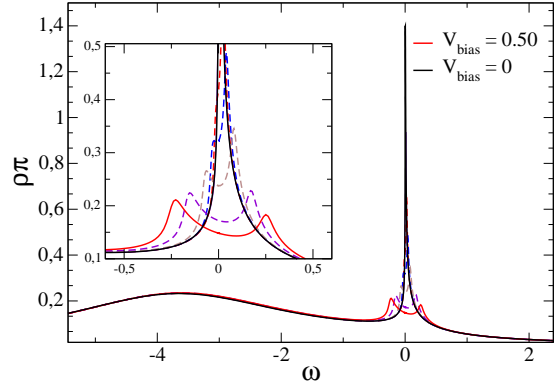}
\caption{\label{fig:fig1} QD spectral density for different applied bias
voltages.}
\end{figure}

For numerical evaluations, we consider a flat conduction band with band width
$2D$ and also take $V_{L}=V_{R}=V$, $\Gamma^{\nu}_{\sigma}=\pi V^2/D$.
As it is clear from Eq. (\ref{eq:current}) and Eq. (\ref{eq:conduct}), the main dependence on 
the conductance at low enough temperatures $T$ is given by the spectral 
weight close to the Fermi level. It is then useful to understand the 
behavior of the spectral density of the QD for different conditions, 
specially in the Kondo regime, where an enhanced conductance is 
expected. In Fig.\ref{fig:fig1} we show the resulting $\rho_{\sigma}$ 
for different applied bias voltages. We take $\Gamma_L = 1$ as our unit and set
$E_d = -4$, $D=10$ and $T=5 \times 10^{-5}$.  We fix the Fermi level $\epsilon_F =0$. 
Note that $\epsilon_F-E_d  \gg \Gamma_L$ so the localized level is always occupied 
($\langle n_{\sigma} \rangle \sim 1$). This corresponds to the Kondo regime, in which
there is a localized spin interacting with the conduction electrons.

At zero bias there are two peaks in the 
spectral density. The one centered close to $E_d$, the charge 
transfer peak, is the result of a non-interacting orbital 
hybridizing with a conduction band. If  temperature were higher than the relevant low-energy scale of the problem, the Kondo temperature $T_K$, 
this would be the only peak in
the spectral 
density. Since $T$ is very low, the low-energy physics is dominated by the Kondo singlet 
between conduction electrons and the localized one. 
The localized spin leads electrons close to 
Fermi level to suffer spin-flip processes giving rise to a screening 
effect. This is the reason for the increase of the spectral weight at 
$\epsilon_F$  shown in the figure, which corresponds to the Kondo peak. 
Its width is related with $T_K$. 
When bias is turned on (dashed curves in Fig. \ref{fig:fig1}) the Fermi level of
each metallic contact is shifted. We set $\mu_L = -\mu_R= eV/2$.
This energy shift produces a splitting of the Kondo 
resonance since conduction electrons coming from both leads contribute to the 
screening process. As a direct consequence the spectral weight at the equilibrium Fermi
level decreases and a lower conductance is expected. In Fig. \ref{fig:fig2}, we
show the differential conductance for several temperatures. Since bias
voltage is applied symmetrically the current $I(V)$ is an odd function and
thus conductance $dI/dV(V)$ an even one. $dI/dV$ curves are
Lorentzian-like with the maximum at zero bias. The value at the maximum and the width
depend on temperature. 

\begin{figure}[t]
\begin{center}
\includegraphics[width=0.43\textwidth]{R2-FIG2}
\caption{\label{fig:fig2} $dI/dV$ vs. $V$ for different temperatures.}
\end{center}
\end{figure}

\begin{figure}[t]
\begin{center}
\includegraphics[width=0.44\textwidth]{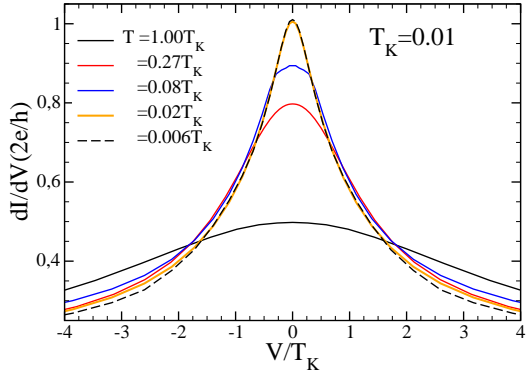}
\caption{\label{fig:fig3} Conductance as a function of temperature obtained 
from lineal response formalism (continuous curve) and differential 
conductance curves evaluated at zero bias(dots) and NRG predictions.}
\end{center}
\end{figure}

From the maximum of the curves of Fig. \ref{fig:fig2}, we can build a 
point-by-point curve of $G$ vs. temperature. The result is shown in 
Fig. \ref{fig:fig3}. We show also a continuous line curve which 
corresponds to an equilibrium calculation of the conductance in the 
linear-response regime by means of Eq. \ref{eq:conduct}. The 
dashed line curve is the empirical formula derived from the NRG calculations\cite{G_E}.  
For high temperatures $T \gg T_K$, there is no Kondo resonance and thus 
the spectral weight at the Fermi level is low. There is an intermediate 
region where thermal fluctuations compete, and at low enough 
temperatures $T \ll T_K$, the Kondo effect is fully developed and 
conductance tends to a saturation value. 
However, at $T \ll T_K$, the NCA overestimates the Friedel's
sum rule and therefore the conductance exceeds the unitary limit.

As it is shown in Fig. \ref{fig:fig3}, there is an excellent agreement between 
the results from the non-equilibrium calculation and the equilibrium ones. 
Moreover, there is also a great 
correspondence to the results from NRG. We stress that the calculation 
of linear response conductance implies only equilibrium quantities 
while the out of equilibrium solving procedure is more complex and 
deals with lesser and greater Green functions. Since the first of this 
approaches is valid just under the condition of proportional couplings, 
the 
agreement we find is a useful check that supports reliability to the 
most general procedure based on the calculation of the current by 
means of Eq. (\ref{eq:current}).

\begin{figure}[t]
\begin{center}
\includegraphics[width=0.45\textwidth]{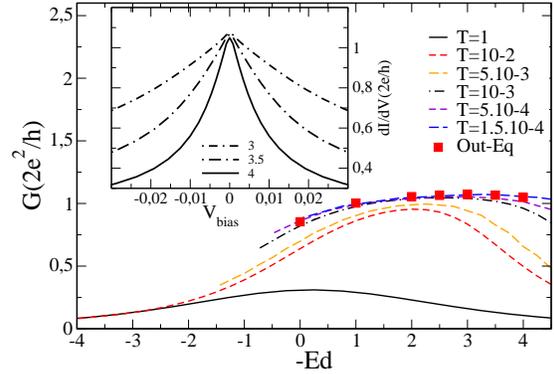}
\caption{\label{fig:fig5} Conductance as a function of gate voltage $V_g$ 
for different temperatures obtained from
lineal response
formalism (continuous curve) and from differential conductance curves evaluated
at zero bias(red dots).}
\end{center}
\end{figure}

We turn now the discussion to the conductance as a function of gate 
voltage $V_g$. The energy of the localized level of the QD is 
proportional to this voltage $eV_g = -E_d$ and thus it is possible to 
perform a transistor-type experiment by the control of this parameter. 
In Fig. \ref{fig:fig5}, we present the NCA results for different 
temperatures. The understanding of this outcome is directly connected 
to our previous analysis of the spectral density. For high temperatures 
(black continuous line in the figure) conductance shows just a 
symmetric peak centered at $V_g=0$. This corresponds to the localized 
level placed at the Fermi energy, the optimum condition for conduction 
electrons to pass from left to right metallic contact. For temperatures 
lower than $T_K$ (dashed curves in the figure) the behavior is completely 
different. As soon as the energy of the localized level is below the 
Fermi energy, the Kondo effect develops and the spectral density shows 
not only the charge transfer peak but also the Kondo resonance. 
This is the reason for the plateau of conductance as a function of 
gate voltage. 
Our results qualitative agree with those obtained previously with NRG  \cite{izu}.
We bring more details on the evolution with temperature.
As it is shown in Fig.\ref{fig:fig5}, at 
finite temperature conductance start to decay at some $V_g$. This 
feature has to do with the fact that $T_K \sim \exp\{E_d\} $, i.e. 
smaller for more negative values of $E_d$. Since temperature is 
finite, $T$ turns bigger than $T_K(E_d)$ at some point destroying the 
Kondo effect and the plateau.
It must be noticed that for $T <T_K$, 
the NCA results are not reliable within the empty orbital regime,
$E_d\geq0$, due to the appearance of a spike with non physical spectral weight at
the Fermi energy. 

For $T=1.5 \times 10^{-4}$ we show (with dots) in Fig. \ref{fig:fig5} the values of 
conductance obtained by the procedure stated previously from the 
calculations out of equilibrium. In the inset of the figure we show
differential conductance curves 
for several $V_g$. We observe that the maximum of the conductance keeps 
the same while the curves get narrower for greater values of gate 
voltage. This is a direct indication of the variation of $T_K$ with $V_g$.        

There are three different way to define the characteristic energy scale $T_K$ from 
the physical magnitudes addressed in this work:  At 
equilibrium it can be obtained from the half-width at half of the 
maximum (HWHM) of spectral density for $T \rightarrow 0$ ($T_K^{\rho}$). Out of 
equilibrium, from the HWHM of differential conductance curves, 
$T_K^{dI/dV}$. From the equilibrium conductance, $T_K^{G}$ is defined as the 
temperature such that $G(T) = G(0)/2$. From our results, $T_K^{G} = 8.2\ 
10^{-3}$, $T_K^{\rho} = 9.1\  10^{-3} = 1.11 T_K^{G}$ and  
$T_K^{dI/dV}=12.1\  10^{-3} = 1.48 T_K^{G}$. As expected, the three values are 
of the same order of magnitude. These realtions can be used to estimate
any one  of them in a case it might not be accessible and to test the model.
%%%%%%%%%%%%%%%%%%%%%%%%%%%%%%%%%%%%%%%%%%%%%%%%%%%%%%%%%%%%%%%%%%%%%%%%
\section{Conclusions}
In this work we study the transport through an interacting quantum dot 
described with the  Anderson model. We use the 
NCA to calculate the conductance as a 
function of bias and gate voltage, and temperature. 
We find a great agreement between the results from the 
non-equilibrium calculations and those from the linear response regime 
which imply only equilibrium quantities. The results for conductance 
versus temperature also agrees with those from NRG calculations. 
We analyze the conductance as a function of gate voltage and observe 
the formation of a plateau for low enough temperatures within the Kondo 
regime in agreement with previous results. As a consequence 
of finite temperature, the conductance decays for a given value of the gate voltage destroying 
the plateau. We finally discuss several methods, in and out of 
equilibrium, which allows the determination of Kondo temperature. We find that the 
values obtained are of the same order and provide numerical relations beetwen them.
While the conductance at equilibrium of the model is known from NRG calculations,
we show that the NCA is able to provide reliable results at a lower computational
cost, and to extend the results out of equilibrium, for which few alternative techniques 
exist. Moreover, the NRG can miss structures in the spectral density which are
not near to the Fermi energy \cite{vau,rou2}. This turned out to be important
to explain a plateau observed in $G(T)$ in $C_{60}$ QDs \cite{roch} on the triplet
side of a quantum phase transition, using the NCA \cite{rou2}.

%%%%%%%%%%%%%%%%%%%%%%%%%%%%%%%%%%%%%%%%%%%%%%%%%%%%%%%%%%%%%%%%%%%%%%%%   
\section{Acknowledgments}
Two of us  (A. A. A. and A. M. L.) are partially supported by CONICET, Argentina.  This work was partially
supported by PIP No 11220080101821 of CONICET, and PICT Nos 2006/483 and
R1776 of the ANPCyT.
%%%%%%%%%%%%%%%%%%%%%%%%%%%%%%%%%%%%%%%%%%%%%%%%%%%%%%%%%%%%%%%%%%%%%%%%

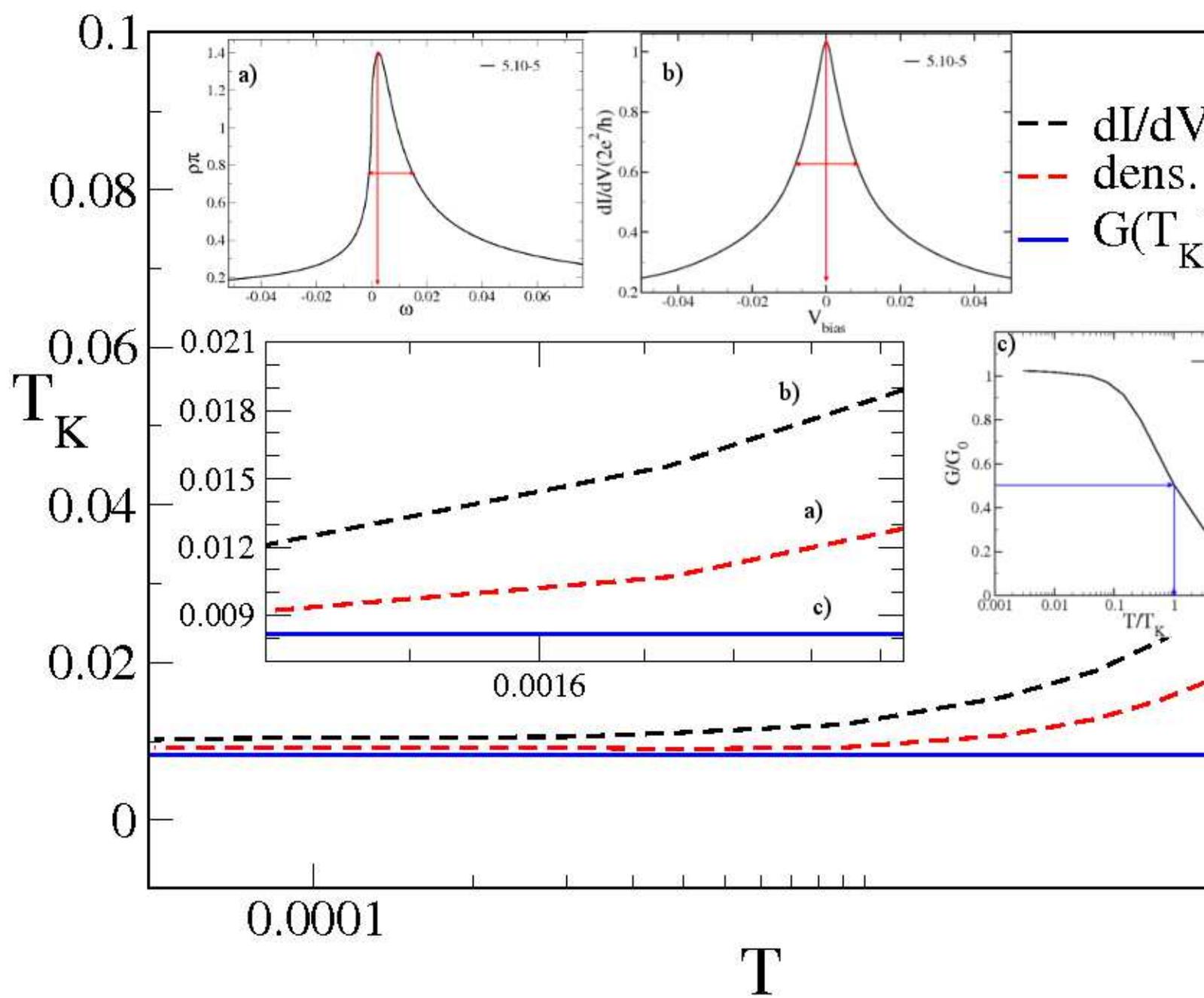
\end{document}